\documentclass[aps,twocolumn,a4paper,floatfix,tightenlines,superscriptaddress,pra,longbibliography]{revtex4-1}
\usepackage[toc,page]{appendix}
\usepackage{braket}
\usepackage{epsfig,graphicx,times}
\usepackage{amstext}
\usepackage{amsmath}
\usepackage{amssymb}
\usepackage{mathrsfs}
\usepackage{dcolumn}
\usepackage{bm}
\usepackage[tight]{subfigure}
\usepackage[colorlinks,linkcolor=red,anchorcolor=blue,citecolor=blue,urlcolor=red]{hyperref}
\usepackage{color}
\usepackage{siunitx}
\usepackage{appendix}
\usepackage{placeins}
\usepackage{textcomp}
\usepackage{bbold}
\usepackage{float}
\usepackage{verbatim}
\usepackage{minitoc}
\usepackage[dvipsnames]{xcolor}

\usepackage{soul}
\usepackage[framemethod=tikz]{mdframed}

\begin{document}

\title{Einstein-Podolsky-Rosen steering of quantum phases \\ in a cavity Bose-Einstein condensate with a single impurity}
%photon blockade in cavity optomechanics with a Bose-Einstein Condensate in the strong coupling regime
%photon blockade with a Bose-Einstein Condensate in strong optomachanics
\author{Shao-Peng Jia}\thanks{Co-first authors with equal contribution}
\affiliation{Key Laboratory of Low-Dimensional Quantum Structures and Quantum Control of Ministry of Education, Department of Physics and Synergetic Innovation Center for Quantum Effects and Applications, Hunan Normal University, Changsha 410081, China}

\author{Baijun Li}\thanks{Co-first authors with equal contribution}
\affiliation{Key Laboratory of Low-Dimensional Quantum Structures and Quantum Control of Ministry of Education, Department of Physics and Synergetic Innovation Center for Quantum Effects and Applications, Hunan Normal University, Changsha 410081, China}

\author{Ya-Feng Jiao}
\affiliation{Key Laboratory of Low-Dimensional Quantum Structures and Quantum Control of Ministry of Education, Department of Physics and Synergetic Innovation Center for Quantum Effects and Applications, Hunan Normal University, Changsha 410081, China}

\author{Hui Jing}\email{jinghui73@gmail.com}
\affiliation{Key Laboratory of Low-Dimensional Quantum Structures and Quantum Control of Ministry of Education, Department of Physics and Synergetic Innovation Center for Quantum Effects and Applications, Hunan Normal University, Changsha 410081, China}

\author{Le-Man Kuang}\email{lmkuang@hunnu.edu.cn}
\affiliation{Key Laboratory of Low-Dimensional Quantum Structures and Quantum Control of Ministry of Education, Department of Physics and Synergetic Innovation Center for Quantum Effects and Applications, Hunan Normal University, Changsha 410081, China}

\date{\today}

\begin{abstract}
We study Einstein-Podolsky-Rosen (EPR) steering properties of quantum phases in the generalized Dicke model (GDM) generated by a cavity Bose-Einstein condensate (BEC) doped with a single impurity. It is shown that the normal  and  superradiant phases of the GDM exhibit much rich EPR steerability.  In the normal phase, there exist  one-way EPR steering  from the cavity field (condensed atoms) to condensed atoms (the cavity field ). In the superradiant phase, there is either one-way or two-way EPR steering for the cavity field and condensed atoms. It is found that the single impurity can act as a single-atom switch of the EPR steering to control  the steering direction of one-way EPR steering and  the transition between one-way and two-way EPR steering through switching on or off the impurity-BEC coupling. It is proved that the EPR steering parameters can witness the quantum phase transition (QPT) occurrence through the sudden change of the EPR steering parameters at the critical point of the QPT.  Our results open a new way to manipulate macroscopic EPR steering and  witness the QPT  between the normal phase and the superradiant phase in the GDM through a microscopic impurity atom.
\end{abstract}

\maketitle

%%%%%%%%%%%%%%%%%%%%%%%%%%%%%%%%%%%%%%%%
\section{\label{level}Introduction}
%%%%%%%%%%%%%%%%%%%%%%%%%%%%%%%%%%%%%%%%%

The Einstein-Podolsky-Rosen (EPR) steering \cite{1,2,3,4,5,6} is an essential feature of non-locality in quantum theory. It can be seen as a quantum correlation or quantum resource   between quantum entanglement and Bell nonlocality since EPR steering requires entanglement as the basic resource to steer the remote state while the correlation implied by EPR steering is not always enough to violate any Bell inequality. EPR steering has a unique asymmetry that is different from quantum entanglement and Bell nonlocality. Due to the asymmetry characteristic, the EPR
steering has potential applications in many aspects such as quantum key distribution  \cite{7,8}, quantum secret sharing \cite{10,11,12,13}, quantum networks \cite{14}, and so on.

In the recent years, much attention has been paid to EPR steering in the community of quantum information and fundamentals of quantum physics.
Continuous and discrete variable EPR steering were also investigated both in theory and in experiment \cite{15,16,17,18,19,20,21,22,23,24,25,26,27,ChenAX,TehRT,Gianani,DaiTZ,ZhuJ,ZhangJ,DesignolleS,SeifertLM,ZhengSS,KongD,HanX,ZengQ,ZhaoSQ,TanHT,ChengWW,PanGZ}. Reid  \cite{24} proposed the criterion to determine continuous variable EPR steering. He and Reid  \cite{19} introduced the criteria for multipartite EPR steering.  Olsen  \cite{25}  showed that one can control the asymmetry of EPR steering by an injected nondegenerate optical parametric oscillator. Bipartite asymmetry EPR steering was also investigated in the cascaded third and fourth-harmonic generation process \cite{26,27} , respectively.
Experimentally, EPR steering has been tested in various physical systems \cite{28,29,31,32,33,34,35,36,37,38}. For instance, Tischler \emph{et al.} \cite{32} gave a conclusive experimental demonstration of one-way steering.  Kunkel \emph{et al.} \cite{33} demonstrated EPR steering between distinct parts of the expanded cloud. Fadel \emph{et al.}  \cite{35}  observed spatial entanglement patterns and EPR steering in Bose-Einstein condensates. Huang  \emph{et al.}  \cite{36} demonstrated that one can generate and control the asymmetry of steady-state EPR steering by engineering asymmetric couplings to the optical field. Li \emph{et al.}  \cite{37} demonstrated quantum steering of orbital angular momentum multiplexed optical fields. Hao \emph{et al.} \cite{38} demonstrated shareability of multipartite EPR Steering. Liu \emph{et al.} \cite{LiuSL} completed experimental demonstration of remotely creating Wigner negativity via quantum steering.

A cavity Bose-Einstein condensate (BEC) \cite{Brennecke2007,Colombe2007} provides an ideal platform for studying strongly-interacting quantum many-body theories, fundamental problems in quantum mechanics, and quantum information processing due to the extremely precise tunability of the interatomic interactions by manipulating $s$-wave scattering length. For example, the Dicke quantum phase transition (QPT) in the cavity-BEC system \cite{Dimer,Baumann1,Yuan2,Keeling,Nagy1,Nagy2,Baumann2,Bastidas,Klinder,Bhaseen,Liu1,Yuan} has been experimentally demonstrated and  extensively studied theoretically. Furthermore, an impurity-doped BEC
 \cite{Ng,Balewski,Schmidt,Johnson,LiQIP,YuanJB,SongYJ,Rammohan1,Rammohan2} builds a micro-macro hybrid quantum systems where the microscopic impurities meet such a macroscopic matter as BEC.
Such a hybrid systems can combine complementary functionalities of micro-macro devices to obtain multitasking capabilities which is crucial for implementing quantum information processing \cite{Schleier-Smith,Kurizkia,Quirion}.

In this paper, we study EPR steering properties of normal and superradiant phases in the generalized Dicke model (GDM) containing a cavity BEC and a single impurity. We show that the normal and superradiant phases in the generalized Dicke model exhibit very rich EPR steerability. We find that the single impurity can act as a single-atom quantum switch of the EPR steering to realize the transition from two-way to unidirectional EPR steering. We also indicate that the EPR steering parameters can witness the QPT occurrence through the sudden change of the EPR steering at the critical point of the QPT.

This paper is structured as follows. In Sec. II, we describe the GDM composed of a cavity BEC doped and a single impurity, and discuss its quantum phase properties. In Sec. III, we investigate the EPR steering properties of the normal phase in the GDM.  In Sec. IV, we study the EPR steering properties of the superradiant phase in the GDM.  In Sec. V, we  show that the EPR steering can witness the  QPT of the GDM. Finally, our conclusions are summarized in Sec. VI.

%%%%%%%%%%%%%%%%%%%%%%%%%%%%%%%%%%%%%%%%%%%%%%%%%%%%%%
\section{\label{level2} Quantum phases in a cavity-BEC with a single impurity}
%%%%%%%%%%%%%%%%%%%%%%%%%%%%%%%%%%%%%%%%%%%%%%%%%%%%%%
We consider an impurity-doped cavity-BEC system, containing a atomic BEC doped a two-level impurity atom (qubit) and a ultrahigh-finesse
optical cavity. The BEC with $N$ identical two-level $^{87}$Rb
atoms is confined in a ultrahigh-finesse optical cavity. The atoms
interact with a single mode cavity filed of frequency $\omega_{c}$ and a
transverse pump field with frequency $\omega_{p}$.
For a large detunings between the cavity field $\omega_{c}$, pump field $\omega_{p}$ and the atomic resonance frequency $\omega_{A}$ (far exceed the rate of atomic spontaneous emission), the atoms only scatter photons either along or transverse to the
cavity axis. Before the pump field turns on, the condensed atoms are
supposed to be prepared in the zero-momentum state
$|p_{x},p_{z}\rangle=|0,0\rangle$. As soon as the drive field is turned on, some
atoms can be excited into nonzero momentum sates $|p_{x},p_{z}\rangle=|\pm
k,\pm k\rangle$ ($\hbar=1$) by the photon scattering of the pump and cavity fields, where $k$ is the wave-vector, which is
approximated equal in the cavity and pump fields.  The zero and nonzero
momentum states i.e., $|0,0\rangle$ and $|\pm k,\pm k\rangle$, are
regarded as two effective levels of the $^{87}$Rb atoms, and can be used to define
 the collective operators of the BEC  $\hat{J_{z}}=\sum_{i}|\pm k,\pm
k\rangle_i \langle \pm k,\pm k|$,
$\hat{J_{+}}=\hat{J_{-}^{\dag}}=\sum_{i} |\pm k,\pm k \rangle_i
\langle 0,0 |$ where $i$ is atoms number index.  In this way, the effective Hamiltonian of the impurity-doped cavity-BEC system is described by the GDM ~\cite{Yuan2}
\begin{eqnarray}
\label{H1}\hat{H}&=&(\omega+\xi_{1}\hat{\sigma}_{z})\hat{a}^{\dag}\hat{a}
+[\omega_{0}-\kappa(\hat{\sigma}_{z}+1)]\hat{J}_{z}
+\frac{\Delta_{Q}}{2}\hat{\sigma}_{z}
 \nonumber\\
&&+\frac{\chi}{N}\hat{J_{z}}^{2}+\frac{\lambda}{\sqrt{N}}\left(\hat{a}+\hat{a}^{\dag}\right)(\hat{J_{+}}+\hat{J_{-}})
\nonumber\\
&&+\xi_{2}\hat{\sigma}_{z}\left(\hat{a}+\hat{a}^{\dag}\right),
\end{eqnarray}
where $\omega$ and $\omega_{0}$ are the effective frequency of the cavity field and the effective transition frequency of the condensed atoms, respectively. $\lambda$ is the coupling strength between the cavity field and the condensed atoms.  $\chi$ denotes the interatomic nonlinear interaction strength. $\kappa$ is the coupling strength between the impurity qubit and the BEC. The other parameters in Eq.(1) are defined by
\begin{eqnarray}
\xi_{1}&=&g_{Q}^{2}/\Delta_{1}, \hspace{0.5cm} \xi_{2}=g_{Q}\Omega_{Q}/\Delta_{1}+g_{Q}\Omega_{Q}/\Delta_{2}, \nonumber\\
\Delta_{Q}&=&\Delta_{2}+g_{Q}^{2}/\Delta_{1}+2\Omega_{Q}^{2}/\Delta_{2},
\end{eqnarray}
where $\Delta_{1}=\omega_{Q}-\omega_{c}$ and  $\Delta_{2}=\omega_{Q}-\omega_{p}$ are the frequency detunings between the
energy separation of the impurity qubit $\omega_{Q}$ and the cavity $\omega_{c}$ and pump
field $\omega_{p}$, respectively. $\hat{\sigma}_{+} (\hat{\sigma}_{-})$
is the raising (lowering) operator of the impurity qubit.  $g_{Q}$ is
the coupling strength between the impurity qubit and the cavity
field, and $\Omega_{Q}$ is the pump Rabi frequency.

The angular momentum operators of the condensed atoms in Eq. (1)  can be expressed  in terms of a single
bosonic mode in the following way, by using the Holstein-Primakoff transformation
\begin{eqnarray}
\hat{J}_{+}&=&\hat{c}^{\dag }\sqrt{N-\hat{c}^{\dag }\hat{c}},
\hspace{0.3cm}
\hat{J}_{-}=\sqrt{N-\hat{c}^{\dag }\hat{c}} \hat{c}, \nonumber\\
\hat{J}_{z}&=&\hat{c}^{\dag}\hat{c}-N/2,
\end{eqnarray}
where the introduced Bose operators obey $[\hat{c}, \hat{c}^{\dag }]=1$.

After taking the average value over a quantum state of the impurity atom and substituting Eq. (3) into the Eq. (1), the Hamiltonian of the impurity-doped Dicke model can be rewritten as
\begin{eqnarray}
\label{H2}\hat{H}^{'}&=&\omega_{1}\hat{a}^{\dag
}\hat{a}+\omega_{2}\hat{c}^{\dag }\hat{c}
+\xi_{2}\delta\left(\hat{a}+\hat{a}^{\dag}\right)+\frac{\chi}{N}(\hat{c}^{\dag }\hat{c})^{2}\nonumber \\
&&+\frac{\lambda}{\sqrt{N}}\left(\hat{a}+\hat{a}^{\dag}\right)\left(\hat{c}^{\dag}\sqrt{N-\hat{c}^{\dag}\hat{c}}
+\sqrt{N-\hat{c}^{\dag}\hat{c}}\hat{c}\right),
\end{eqnarray}
where a constant term has been ignored which does not affect the dynamic evolution of the system. The two effective frequencies are given by
\begin{equation}
\omega_1=\omega +\xi _{1}\delta,\hspace{1cm}\omega_{2}={\omega-\chi^{''}-\kappa(1+ \delta)},
\end{equation}
where $\chi^{\prime\prime}$ is a nonlinear parameter which depends on the $s$-wave scattering lengths of the condensed atoms, and $\delta=\langle \hat{\sigma}_{z}\rangle$ is the impurity population ($-1\leq \delta \leq 1$).

The QPT for the impurity-doped Dicke model [Eq. (4)] can be investigated in terms of the mean-field approach. By translating the bosonic modes in the following way
\begin{equation}
\hat{d}=\hat{a}+\sqrt{N}\alpha, \hspace{0.5cm}
\hat{b}=\hat{c}-\sqrt{N}\beta,
\end{equation}
where $\alpha$ and $\beta$ are real numbers, they describe macroscopic excitation population of the cavity field and the condensed atoms, respectively. In the normal phase, there is no macroscopic excitation population in the cavity and condensed atoms modes, $\alpha=\beta=0$.  In the super-radiant phase, both modes acquire non-zero macroscopic excitation population, hence we have $\alpha\neq 0$ and $\beta\neq 0$.
The displaced bosonic operators of the cavity field and condensed atoms obey  $[\hat{d}, \hat{d}^{\dag }]=1$ and  $[\hat{b}, \hat{b}^{\dag }]=1$.

When the cavity-atom coupling strength
$\lambda$ is chosen as  the QPT parameter, the critical coupling strength $\lambda_c$ at the QPT point is given by
\begin{equation}
\label{lac} \lambda_{c}= \frac{1}{2}\sqrt{(\omega +\xi _{1}\delta)[\omega-\kappa(1+ \delta)]},
\end{equation}
which indicates that  the critical coupling strength $\lambda_{c}$
can be manipulated continuously through  changing the impurity population $\delta$.

When $\lambda< \lambda_{c}$, the cavity-BEC system is in the normal phase in which there is no excited population of the cavity field and the condensed atoms.
When $\lambda> \lambda_{c}$, the cavity-BEC system is in the super-radiant phase in which both of the cavity field and the condensed atoms have excited population.  In the normal phase, the Hamiltonian of the cavity-BEC system is given by
\begin{eqnarray}
\hat{H}_{N}&=&\omega_1\hat{d}^{\dagger }\hat{d}+\omega_2\hat{b}^{\dagger }\hat{b}
+\lambda(\hat{d}+\hat{d}^{\dagger })(\hat{b}+\hat{b}^{\dagger }) \nonumber\\
& &+\xi_{2}\delta (\hat{d}+\hat{d}^{\dagger }),
\end{eqnarray}
where $\hat{b}^{\dagger} (\hat{b})$ is the creation (annihilation)  operator of the condensed atoms introduced by the  Holstein-Primakoff transformation

In the superradiant phase, the Hamiltonian of the cavity-BEC system is given by
\begin{eqnarray}
\hat{H}_{S}&=&\omega_1\hat{d}^{\dagger }\hat{d}+\omega_{3}\hat{b}^{\dagger }\hat{b}
+\zeta (\hat{d}+\hat{d}^{\dagger })(\hat{b}+\hat{b}^{\dagger })\nonumber\\
& &+\eta (\hat{b}
+\hat{b}^{\dagger })^{2}+\xi _{2}\delta (\hat{d}+\hat{d}^{\dagger }),
\end{eqnarray}
where the effective frequency of the condensed atoms and the effective coupling strength between the cavity field and the condensed atoms are given  by
\begin{eqnarray}
\omega_{3}&=&\omega _{r}-\chi ^{^{\prime \prime }}-\kappa (1+\delta )+2\chi \beta ^{2}+\frac{\lambda \alpha \beta }{K}, \nonumber\\
\zeta&=&\lambda(K-\frac{\beta ^{2}}{K}), \hspace{0.2cm} \eta=\chi \beta ^{2}+\frac{\lambda \alpha \beta (2+\beta ^{2})}{2K^{3}},
\end{eqnarray}
where $K=\sqrt{1-\beta^2}$. The excitation populations of the cavity field $\alpha$ and the condensed atoms $\beta$ in the super-radiant phase are
given by
\begin{eqnarray}
\alpha^{2}&=&\frac{\lambda^{2}(4\lambda^{2}-\omega_{1}\omega_{2})(4\lambda^{2}+\omega_{1}\omega_{2}+2\chi \omega_{1})}{\omega_{1}^{2}(\chi \omega_{1}+4\lambda^{2})^{2}}, \nonumber\\
\beta ^{2}&=&\frac{4\lambda^{2}-\omega_{1}\omega_{2}}{2\chi \omega_{1}+8\lambda^{2}}.
\end{eqnarray}

In the following sections we will investigate quantum steering properties in the two quantum phases of the impurity-doped Dicke model in terms of the two Hamiltonian given by Eqs. (8) and (9).

%%%%%%%%%%%%%%%%%%%%%%%%%%%%%%%%%%%%%%%%%%
\section{\label{level2} EPR steering between photons and atoms in the normal phase}
%%%%%%%%%%%%%%%%%%%%%%%%%%%%%%%%%%%%%%%%%%

In this section, we study  properties of quantum steering between photons in the cavity field and condensed atoms in the normal phase. We will find the ground-state wave function of the normal phase of  the impurity-doped Dicke model,  calculate the quantum steering parameters and discuss quantum steering characteristics of the normal phase.

The Hamiltonian of  the normal phase  is given by Eq. (8). In order to solve this Hamiltonian, we move to a position-momentum representation
\begin{eqnarray}
\hat{x}_{1}&=&\frac{1}{\sqrt{2\omega_{1}}}(\hat{d}^{\dagger}+\hat{d}),\hspace{0.2cm}
\hat{p}_{1}=i\sqrt{\frac{\omega_{1}}{2}}(\hat{d}^{\dagger}-\hat{d}) ,\notag\\
\hat{x}_{2}&=&\frac{1}{\sqrt{2\omega_{2}}}(\hat{b}^{\dagger}+\hat{b}),\hspace{0.2cm}
\hat{p}_{2}=i\sqrt{\frac{\omega_{2}}{2}}(\hat{b}^{\dagger}-\hat{b}).
\end{eqnarray}

Substituting Eq. (12) into the Eq. (8), we can obtain the normal phase Hamiltonian in the position-momentum representation
\begin{align}
\hat{H}_{N}=&\frac{\hat{p}_{1}^{2}}{2}+\frac{1}{2}\omega_{1}^{2}\hat{x}_{1}^{2}+\frac{%
\hat{p}_{2}^{2}}{2}+\frac{1}{2}\omega_{2}^{2}\hat{x}_{2}^{2}\notag\\
&+2\lambda \sqrt{\omega_{1}\omega_{2}}\hat{x}_{1}\hat{x}_{2}  +\xi _{2}\delta \sqrt{2\omega_{1}}\hat{x}_{1},
\end{align}
where the last term indicates that the cooperative effect, which can produce an effective driving on the cavity field and can be manipulated through changing the impurity population $\delta$.

In order to diagonalize the  normal phase Hamiltonian [Eq. (13)], we introduce two super-mode oscillators with the following position-momentum representation
\begin{align}
&\hat{Q}_{1}=\hat{x}_{1}\cos \theta -\hat{x}_{2}\sin \theta,\hspace{0.2cm}
\hat{P}_{1}=\hat{p}_{1}\cos \theta -\hat{p}_{2}\sin \theta,\notag\\
&\hat{Q}_{2}=\hat{x}_{1}\sin \theta +\hat{x}_{2}\cos \theta,\hspace{0.2cm}
\hat{P}_{2}=\hat{p}_{1}\sin \theta +\hat{p}_{2}\cos \theta,
\end{align}
which lead to the following diagonalized Hamiltonian
\begin{align}
\hat{H}_{N}=&\frac{1}{2}\hat{P}_{1}^{2}+\frac{1}{2}\Omega_{1}^{2}\hat{Q}_{1}^{2}+\frac{1}{2}\hat{P}_{2}^{2}+
\frac{1}{2}\Omega_{2}^{2}\hat{Q}_{2}^{2}\notag\\%
&+\xi _{2}\delta\sqrt{2\omega_{1}}\left(\hat{Q}_{1}\cos\theta+\hat{Q}_{2}\sin\theta\right),
\end{align}
where the two super-mode frequencies and the mixing angle are given by
\begin{align}
&\Omega_{1,2}^{2}
=\frac{1}{2}\left[\omega_{1}^{2}+\omega_{2}^{2}\pm\sqrt{\left(\omega_{1}^{2}-\omega_{2}^{2}\right)^{2}+16\lambda^{2}\omega_{1}\omega_{2}}\right], \\\
&\tan 2\theta=\frac{4\lambda\sqrt{\omega_{1}\omega_{2}}}{\omega_{2}^{2}-\omega_{1}^{2}}.
\end{align}
The Eq. (15) indicates that the normal-phase Hamiltonian can act as two driving super-mode oscillators.

In order to obtain the eigenstates of the normal-phase Hamiltonian, we use the displacing transformation for the position operators of the two super-mode oscillators
\begin{align}
&\hat{Q}_{1}^{'}=\hat{Q}_{1}+\frac{\xi _{2}\delta\sqrt{2\omega_{1}}\cos\theta}{\Omega_{1}^{2}}, \hspace{0.2cm}
\hat{Q}_{2}^{'}=\hat{Q}_{2}+\frac{\xi _{2}\delta\sqrt{2\omega_{1}}\sin\theta}{\Omega_{2}^{2}}.
\end{align}
Then we can rewrite the normal-phase Hamiltonian as
\begin{align}
\hat{H}_{N}=&\frac{1}{2}\hat{P}_{1}^{2}+\frac{1}{2}\Omega_{1}^{2}\hat{Q}_{1}^{'2}+
\frac{1}{2}\hat{P}_{2}^{2}+\frac{1}{2}\Omega_{2}^{2}\hat{Q}_{2}^{'2},
\end{align}
where the constants have been neglected which do not affect the dynamic evolution of the system.

Obviously,  the eigenstates of above normal-phase Hamiltonian are number states, i.e.,
\begin{align}
|n,m\rangle_{\hat{A}_{1}^{'},\hat{A}_{2}^{'}}=\frac{\hat{A}_{1}^{'\dagger n}\hat{A}_{2}^{'\dagger m}}{\sqrt{n!m!}} |0,0\rangle_{\hat{A}_{1}^{'}
\hat{A}_{2}^{'}},\hspace{0.5cm}n,m=0,1,2, \cdots
\end{align}
where the annihilation and creation operators of the super-modes are defined by
\begin{align}
\hat{Q}_{1}^{'}=\frac{1}{\sqrt{2\Omega_{1}}}\left(\hat{A}_{1}^{'\dagger}+\hat{A}_{1}^{'}\right),\hspace{0.2cm}
\hat{P}_{1}=i\sqrt{\frac{\Omega_{1}}{2}}\left(\hat{A}_{1}^{'\dagger}-\hat{A}_{1}^{'}\right), \notag\\
\hat{Q}_{2}^{'}=\frac{1}{\sqrt{2\Omega_{2}}}\left(\hat{A}_{2}^{'\dagger}+\hat{A}_{2}^{'}\right),\hspace{0.2cm}
\hat{P}_{2}=i\sqrt{\frac{\Omega_{2}}{2}}\left(\hat{A}_{2}^{'\dagger}-\hat{A}_{2}^{'}\right).
\end{align}

From Eq. (20) we can obtain the ground-state wave function of the normal phase in the super-mode representation
\begin{align}
\Psi_{0}(\hat{Q}_{1}^{'},\hat{Q}_{2}^{'})=\left(\frac{\Omega_{1}\Omega_{2}}{\pi^{2}}\right)^{\frac{1}{4}}\exp\left[-\frac{1}{2}\left(\Omega_{1}\hat{Q}_{1}^{'2}
+\Omega_{2}\hat{Q}_{2}^{'2}\right)\right],
\end{align}
which indicates that the ground-state wave function of the normal phase is unentangled in the super-mode representation. Substituting the Eqs. (14) and (1) into Eq. (22), we can obtain the ground-state wave function of the normal phase in the original position-momentum representation
\begin{align}
\Psi_{0}(\hat{x}_{1},\hat{x}_{2})=&\left(\frac{\Omega_{1}\Omega_{2}}{\pi^{2}}\right)^{\frac{1}{4}}\notag\\
&\times \exp\left[-\frac{1}{2}
\Omega_{1}\left((\hat{x}_{1}+x_{01})\cos \theta -\hat{x}_{2}\sin \theta \right)^{2}\right]\notag\\
&\times\exp\left[-\frac{1}{2}
\Omega_{2}\left((\hat{x}_{1}+x_{02})\sin \theta +\hat{x}_{2}\cos \theta\right)^{2}\right],
\end{align}
where we have introduced the following parameters
\begin{equation}
x_{01}=\frac{\xi _{2}\delta\sqrt{2\omega_{1}}}{\Omega_{1}^{2}}, \hspace{0.2cm} x_{02}=\frac{\xi _{2}\delta\sqrt{2\omega_{1}}}{\Omega_{2}^{2}}.
\end{equation}

\begin{figure}[t]
  \centerline{
  \includegraphics[width=0.5\textwidth]{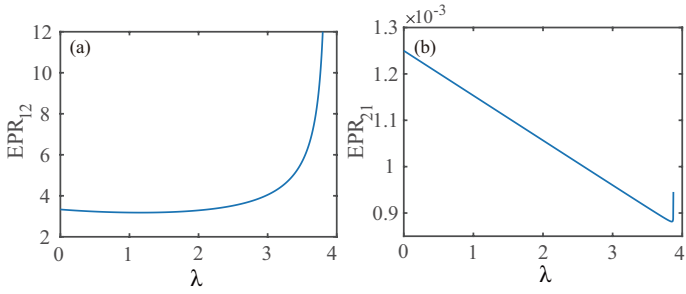}}
  \caption{EPR steering with the impurity atom versus the coupling strength $\lambda$ between the cavity field and the condensed atoms, when the mixing angle obeys the condition  $\sin 2\theta >0$. The numerical simulation parameters are $\omega=400\omega_{r},\ \chi^{"}=0.1\omega_{r},\ \kappa=0.5\omega_{r},\ \xi_{1}=0.001\omega_{r}$, and $\delta=0.5\omega_{r}$.} \label{fig5}
\end{figure}

From Eq. (24) we can see that the ground-state wave function of the normal phase  is an entangled state in the original position-momentum representation. In what follows we study EPR steering properties of the ground-state wave function. The ability of EPR steering for a given two-mode quantum state can be measured by the two steering parameters,
\begin{align}
E_{12}=\left(\Delta \hat{x}_{1}\right)^{2}-\frac{\langle \hat{x}_{1},\hat{x}_{2} \rangle}{\left(\Delta \hat{x}_{2}\right)^{2}} < \frac{1}{2},\notag\\
E_{21}=\left(\Delta\hat{x}_{2}\right)^{2}-\frac{\langle\hat{x}_{1},\hat{x}_{2}\rangle}{\left(\Delta\hat{x}_{1}\right)^{2}} < \frac{1}{2},
\end{align}
where the cross-correlation function of  the cavity mode and the condensed-atom mode is defined by
\begin{align}
\langle \hat{x}_{1},\hat{x}_{2}\rangle&=\frac{1}{2}(\langle \hat{x}_{1}\hat{x}_{2}\rangle+\langle \hat{x}_{2}\hat{x}_{1}\rangle)-\langle \hat{x}_{1}\rangle\langle \hat{x}_{2}\rangle.
\end{align}

The first inequality $E_{12} < 1/2$ means the steerability from  the condensed-atom mode (atoms) to the cavity-field mode (photons) while  the second inequality  $E_{21} < 1/2$ indicates the steering from the cavity-field mode (photons) to the condensed-atom mode (atoms).

For the ground-state wave function of the normal phase [Eq. (23)], we can obtain the variances of the cavity-field mode and the condensed-atom mode,
\begin{align}
(\Delta \hat{x}_{1})^{2}&=\frac{\Omega_{2}\cos\theta^{2}+\Omega_{1}\sin\theta^{2}}{2\Omega_{1}\Omega_{2}},\\
(\Delta \hat{x}_{2})^{2}&=\frac{\Omega_{2}\sin^{2}\theta+\Omega_{1}\cos^{2}\theta
}{2\Omega_{1}\Omega_{2}},
\end{align}
where the cross-correlation function of  the cavity-field mode and the condensed-atom mode is given by
\begin{equation}
\langle \hat{x}_{1},\hat{x}_{2}\rangle=\frac{\Omega_{1}-\Omega_{2}}{4\Omega_{1}\Omega_{2}}\sin2\theta.
\end{equation}

Substituting Eqs. (27), (28) and (29) into Eq. (25), we can obtain the two EPR steering parameters
\begin{equation}
E_{12}=\frac{\sin2\theta(\Omega_{1}-\Omega_{2})[\sin2\theta(\Omega_{1}-\Omega_{2}) -4\Omega_{1}\Omega_{2}]+4\Omega_{1}\Omega_{2}}{4\Omega_{1}\Omega_{2}[(\Omega_{1}+\Omega_{2})
+\cos2\theta(\Omega_{1}-\Omega_{2})]},
\end{equation}
\begin{equation}
E_{21}=\frac{\sin2\theta(\Omega_{1}-\Omega_{2})[\sin2\theta(\Omega_{1}-\Omega_{2}) -4\Omega_{1}\Omega_{2}]+4\Omega_{1}\Omega_{2}}{4\Omega_{1}\Omega_{2}[(\Omega_{1}+\Omega_{2})
-\cos2\theta(\Omega_{1}-\Omega_{2})]},
\end{equation}
where the $\sin2\theta$ and $\cos2\theta$ can be obtained by the Eq. (17).

When we consider the case of $\omega_{1}> \omega_{2}$ and $\tan2\theta<0$, we have
\begin{align}
\sin2\theta=\pm\sqrt{\frac{16\lambda^{2}\omega_{1}\omega_{2}}{(\omega_{1}^{2}-\omega_{2}^{2})^{2}+16\lambda^{2}\omega_{1}\omega_{2}}}, \\
\cos2\theta=\mp\sqrt{\frac{(\omega_{1}^{2}-\omega_{2}^{2})^{2}}{(\omega_{1}^{2}-\omega_{2}^{2})^{2}+16\lambda^{2}\omega_{1}\omega_{2}}}.
\end{align}

%%%%%%%%%%%%%%%%%%%%%%%%%%%%%%%%%%%%%%%%%%%%%%%%%%%%%%%%%%%%%%%%%%%%%%
%In what follows we consider the case of $\omega_{1}<\omega_{2}$. In this case $\tan2\theta>0$, we have
%\begin{align}
%\sin2\theta=\pm\sqrt{\frac{16\lambda^{2}\omega_{1}\omega_{2}}{(\omega_{1}^{2}-\omega_{2}^{2})^{2}+16\lambda^{2}\omega_{1}\omega_{2}}}\notag\\
%\cos2\theta=\pm\sqrt{\frac{(\omega_{2}^{2}-\omega_{1}^{2})^{2}}{(\omega_{1}^{2}-\omega_{2}^{2})^{2}+16\lambda^{2}\omega_{1}\omega_{2}}}
%\end{align}
%%%%%%%%%%%%%%%%%%%%%%%%%%%%%%%%%%%%%%%%%%%%%%%%%

\begin{figure}[t]
  \centerline{
  \includegraphics[width=0.5\textwidth]{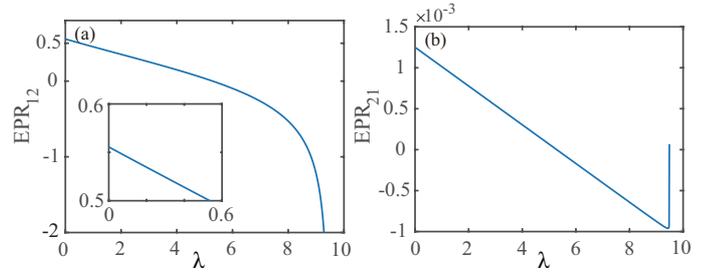}}
  \caption{EPR steering without the impurity atom versus the coupling strength $\lambda$ between the cavity field and the condensed atoms, when the mixing angle obeys the condition $\sin 2\theta >0$. The numerical simulation parameters are $\omega=400\omega_{r}$ and $\chi^{"}=0.1\omega_{r}$.} \label{fig5}
\end{figure}

In what follows we  numerically analyze the EPR steering properties of the ground-state wave function of the normal phase in terms of the analytical expressions in Eqs. (30) and (31) with $\omega_1>\omega_2,\ \sin 2\theta>0$, and $\cos 2\theta<0$. In Fig. 1, we plot the two EPR parameters in the normal phase as a function of the coupling strength between the cavity field and condensed atoms in the presence of the impurity atom with the mixing angle obeys the condition  $\sin 2\theta >0$. Here we take the $\omega_r$ as the scale frequency. The numerical simulation parameters are chosen as $\omega=400\omega_{r},\ \chi^{"}=0.1\omega_{r},\ \kappa=0.5\omega_{r},\ \xi_{1}=0.001\omega_{r}$, and $\delta=0.5\omega_{r}$.
Fig. 1 (b) shows that in the normal phase, when the mixing angle obeys the condition $\sin 2\theta >0$, the cavity field can unidirectionally steer the condensed atoms with the impurity atoms over the whole regime of the field-BEC coupling strength. However, Fig. 1 (a) indicates that the condensed atoms can not steer the cavity field in the whole regime of the field-BEC coupling strength in the normal phase.

\begin{figure}[t]
  \centerline{
  \includegraphics[width=0.5\textwidth]{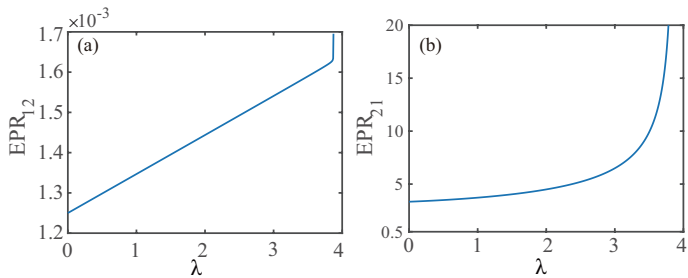}}
  \caption{EPR steering with the impurity atom versus the coupling strength $\lambda$ between the cavity field and the condensed atoms, when
the mixing angle obeys the condition $\sin 2\theta <0$. The numerical simulation parameters are $\omega=400\omega_{r},\ \chi^{"}=0.1\omega_{r},\ \kappa=0.5\omega_{r},\ \xi_{1}=0.001\omega_{r}$, and $\delta=0.5\omega_{r}$.} \label{fig5}
\end{figure}

In Fig. 2 we plot the two EPR parameters in the normal phase as the function of the coupling strength between the cavity field and condensed atoms without the impurity atom, when the mixing angle obeys the condition  $\sin 2\theta >0$. We take the related parameters as $\omega=400\omega_{r}$ and $\chi^{"}=0.1\omega_{r}$. From Fig. 2 (a) and (b) we can see that   $E_{12}<0.5$ and   $E_{21}< 0.5$  in the regime of  $0.53<\lambda<\lambda_c$. This means that the exists the  two-way EPR steering between the cavity field and condensed atoms in the coupling regime of $0.53<\lambda<\lambda_c$.  However, in the small regime of the field-BEC coupling strength,  $0<\lambda<0.53$, we find $E_{12}>0.5$ and   $E_{21}< 0.5$, shown in the inset in Fig. 2 (a). Hence, there exists the  one-way EPR steering from the cavity field to the condensed atoms in the small regime of the field-BEC coupling strength.

Combining the EPR steering results from Fig. 1 (with the impurity atom) and Fig. 2 (without the impurity atom), we can find that the impurity atom can induce the transition from the two-way EPR steering between the cavity field and condensed atoms to the one-way EPR steering. Thus the impurity atom can be regarded as a single atom switch that realizes the transition from the two-way EPR steering to the one-way EPR steering.

\begin{figure}[t]
  \centerline{
  \includegraphics[width=0.5\textwidth]{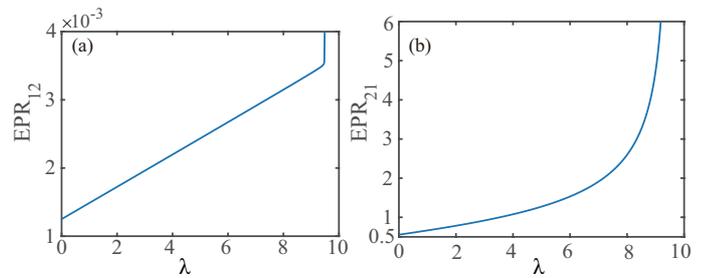}}
  \caption{EPR steering without the impurity atom versus the coupling strength $\lambda$ between the cavity field and the condensed atoms, when
the mixing angle obeys the condition $\sin 2\theta <0$. The numerical simulation parameters are $\omega=400\omega_{r}$ and $\chi^{"}=0.1\omega_{r}$.} \label{fig5}
\end{figure}

In Fig. 3 we plot the two EPR parameters in the normal phase as the function of the coupling strength between the cavity field and condensed atoms in the presence of the impurity atom, when the mixing angle obeys the condition  $\sin 2\theta <0$. Here the related parameters are chosen as $\omega=400\omega_{r},\ \chi^{"}=0.1\omega_{r},\ \kappa=0.5\omega_{r},\ \xi_{1}=0.001\omega_{r}$, and $\delta=0.5\omega_{r}$. Fig. 3 (a) indicates that in the normal phase, the condensed atoms can unidirectionally steer the cavity field in the whole regime of $\lambda<\lambda_c$ with $\lambda_c =3.87\omega_{r}$  when the mixing angle obeys the condition  $\sin 2\theta <0$. However, we find that the cavity field can not steer the condensed atoms in the whole coupling regime of  $\lambda<\lambda_c$  in the normal phase, shown in Fig. 3 (b).

In Fig. 4 we plot the two EPR parameters in the normal phase as the function of the coupling strength between the cavity field and condensed atoms without the presence of the impurity atom, when the mixing angle obeys the condition  $\sin 2\theta < 0$. We take the related parameters as $\omega=400\omega_{r}$ and $\chi^{"}=0.1\omega_{r}$. From Figs. 4 (a)  and (b) we can see that   $E_{12}<0.5$ and   $E_{21}> 0.5$  in the whole coupling regime of the normal phase.  This means that the condensed atoms can steer  the cavity field while  the cavity field can not steer the condensed atoms.

Comparing the EPR steering results of Fig.3 with Fig. 1, it is straightforward to see that one can change the direction of the one-way EPR steering through adjusting the sign of the mixing angle function $\sin 2\theta$. When $\sin 2\theta>0$, the cavity field can unidirectionally steer the condensed atoms while when $\sin 2\theta<0$, the condensed atoms can unidirectionally steer the cavity field.

From the abve numerical analysis we can conclude that in the normal phase, there exist either one-way EPR steering from the cavity field to condensed atoms or one-way EPR steering from condensed atoms to the cavity field, and the direction of the EPR steering can be controlled through adjusting the mixing-angle function $\sin 2\theta$.  However,  there is not two-way EPR steering between the cavity field and condensed atoms in the normal phase.
The single impurity can be regarded as a single-atom switch  to control the EPR steering direction. The one-way EPR steering can be changed to the two-way EPR steering through switching off the impurity-BEC coupling while the two-way EPR steering can be turned to the one-way EPR steering through switching on the impurity-BEC coupling.

%%%%%%%%%%%%%%%%%%%%%%%%%%%%%%%%%%%%%%%%%%
\section{\label{level2} EPR steering between photons and atoms in the superadiant phase}
%%%%%%%%%%%%%%%%%%%%%%%%%%%%%%%%%%%%%%%%%%

In this section, we study  properties of quantum steering between photons in the cavity field and condensed atoms in the superradiant phase.
The Hamiltonian of  the superradiant  phase is given by Eq. (9). In order to solve this Hamiltonian, we move to a position-momentum representation
\begin{eqnarray}
\hat{x}_{1}&=&\frac{1}{\sqrt{2\omega_{1}}}(\hat{d}^{\dagger}+\hat{d}),\hspace{0.2cm}
\hat{p}_{1}=i\sqrt{\frac{\omega_{1}}{2}}(\hat{d}^{\dagger}-\hat{d}) ,\notag\\
\hat{x}_{3}&=&\frac{1}{\sqrt{2\omega_{3}}}(\hat{b}^{\dagger}+\hat{b}),\hspace{0.2cm}
\hat{p}_{3}=i\sqrt{\frac{\omega_{3}}{2}}(\hat{b}^{\dagger}-\hat{b}).
\end{eqnarray}

Substituting Eq. (34) into the Eq. (9), we can obtain the superadiant phase Hamiltonian in the position-momentum representation
\begin{align}
\hat{H}_{S}=&\frac{\hat{p}_{1}^{2}}{2}+\frac{1}{2}\omega_{1}^{2}\hat{x}_{1}^{2}+\frac{%
\hat{p}_{3}^{2}}{2}+\frac{1}{2}\omega_{3}(\omega_{3} + 4\eta)\hat{x}_{3}^{2}\notag\\
&+2\zeta\sqrt{2\omega_{1}\omega_3}\hat{x}_{1}\hat{x}_{3}  +\xi_2\delta \sqrt{2\omega_{1}}\hat{x}_{1}.
\end{align}
Similar to the case of the normal phase, the cooperative influence can cause an  effective driving on the cavity field, which can be manipulated through changing the impurity population $\delta$.

In order to diagonalize the superradiant phase Hamiltonian [Eq. (35)], we introduce two super-mode oscillators of the superradiant phase with the following position-momentum representation
\begin{align}
&\hat{Q}_{1}=\hat{x}_{1}\cos \theta -\hat{x}_{3}\sin \theta,\hspace{0.2cm}
\hat{P}_{1}=\hat{p}_{1}\cos \theta -\hat{p}_{3}\sin \theta,\notag\\
&\hat{Q}_{3}=\hat{x}_{1}\sin \theta +\hat{x}_{3}\cos \theta,\hspace{0.2cm}
\hat{P}_{3}=\hat{p}_{1}\sin \theta +\hat{p}_{3}\cos \theta,
\end{align}
which lead to the following diagonalized Hamiltonian
\begin{align}
\hat{H}_{S}=&\frac{1}{2}\hat{P}_{1}^{2}+\frac{1}{2}\Omega_{1}^{2}\hat{Q}_{1}^{2}+\frac{1}{2}\hat{P}_{3}^{2}+
\frac{1}{2}\Omega_{2}^{2}\hat{Q}_{3}^{2}\notag\\%
&+\xi _{2}\delta\sqrt{2\omega_{1}}\left(\hat{Q}_{1}\cos\theta+\hat{Q}_{3}\sin\theta\right),
\end{align}
where the two super-mode frequencies and the mixing angle are given by
\begin{eqnarray}
\Omega_{1,3}^{2}&=&\frac{1}{2}\left[\omega_{1}^{2}+\omega_{3}^{2}+4\eta\omega_3 \right. \notag\\
& &\left.\pm\sqrt{\left(\omega_{1}^{2}-\omega_{3}^{2}-4\eta\omega_3\right)^{2}+16\lambda^{2}\omega_{1}\omega_{3}}\right], \\\
\tan 2\theta&=&\frac{4\lambda\sqrt{\omega_{1}\omega_{3}}}{\omega_{3}^{2}+4\eta\omega_3-\omega_{1}^{2}}.
\end{eqnarray}
The Hamiltonian (37) indicates that the superradiant-phase Hamiltonian can act as two driving super-mode oscillators.

In order to obtain the eigenstates of the superradiant-phase Hamiltonian, we make the displacing transformation for the position operators of the two super-mode oscillators
\begin{align}
&\hat{Q}_{1}^{'}=\hat{Q}_{1}+\frac{\xi _{2}\delta\sqrt{2\omega_{1}}\cos\theta}{\Omega_{1}^{2}}, \hspace{0.2cm}
\hat{Q}_{3}^{'}=\hat{Q}_{3}+\frac{\xi _{2}\delta\sqrt{2\omega_{1}}\sin\theta}{\Omega_{3}^{2}}.
\end{align}
Then we can rewrite the superradiant-phase Hamiltonian (37) as
\begin{align}
\hat{H}_{S}=&\frac{1}{2}\hat{P}_{1}^{2}+\frac{1}{2}\Omega_{1}^{2}\hat{Q}_{1}^{'2}+
\frac{1}{2}\hat{P}_{3}^{2}+\frac{1}{2}\Omega_{3}^{2}\hat{Q}_{3}^{'2},
\end{align}
where the constants have been neglected which do not affect the dynamic evolution of the system.

Obviously,  the eigenstates of above superradiant-phase Hamiltonian are number states, i.e.,
\begin{align}
|n,m\rangle_{\hat{A}_{1}^{'},\hat{A}_{3}^{'}}=\frac{\hat{A}_{1}^{'\dagger n}\hat{A}_{3}^{'\dagger m}}{\sqrt{n!m!}} |0,0\rangle_{\hat{A}_{1}^{'}
\hat{A}_{3}^{'}},\hspace{0.5cm}n,m=0,1,2, \cdots
\end{align}
where the annihilation and creation operators of the super-modes are defined by
\begin{align}
\hat{Q}_{1}^{'}=\frac{1}{\sqrt{2\Omega_{1}}}\left(\hat{A}_{1}^{'\dagger}+\hat{A}_{1}^{'}\right),\hspace{0.2cm}
\hat{P}_{1}=i\sqrt{\frac{\Omega_{1}}{2}}\left(\hat{A}_{1}^{'\dagger}-\hat{A}_{1}^{'}\right), \notag\\
\hat{Q}_{3}^{'}=\frac{1}{\sqrt{2\Omega_{3}}}\left(\hat{A}_{3}^{'\dagger}+\hat{A}_{3}^{'}\right),\hspace{0.2cm}
\hat{P}_{3}=i\sqrt{\frac{\Omega_{3}}{2}}\left(\hat{A}_{3}^{'\dagger}-\hat{A}_{3}^{'}\right).
\end{align}

From Eq. (42) we can obtain the ground-state wave function of the superradiant phase in the super-mode representation
\begin{align}
\Psi_{0}(\hat{Q}_{1}^{'},\hat{Q}_{3}^{'})=\left(\frac{\Omega_{1}\Omega_{3}}{\pi^{2}}\right)^{\frac{1}{4}}\exp\left[-\frac{1}{2}\left(\Omega_{1}\hat{Q}_{1}^{'2}
+\Omega_{3}\hat{Q}_{3}^{'2}\right)\right],
\end{align}
which indicates that the ground-state wave function of the superradiant phase is unentangled in the super-mode representation. Substituting Eqs. (36) and (40) into Eq. (44), we can obtain the ground-state wave function of the superradiant phase in the original position-momentum representation
\begin{align}
\Psi_{0}(\hat{x}_{1},\hat{x}_{3})=&\left(\frac{\Omega_{1}\Omega_{3}}{\pi^{2}}\right)^{\frac{1}{4}}\notag\\
&\times \exp\left[-\frac{1}{2}
\Omega_{1}\left((\hat{x}_{1}+x_{01})\cos \theta -\hat{x}_{3}\sin \theta \right)^{2}\right]\notag\\
&\times\exp\left[-\frac{1}{2}
\Omega_{3}\left((\hat{x}_{1}+x_{03})\sin \theta +\hat{x}_{3}\cos \theta\right)^{2}\right],
\end{align}
where $x_{01}$ is given by Eq. (24) and the parameter $x_{03}$ is defined by
\begin{equation}
 x_{03}=\frac{\xi _{2}\delta\sqrt{2\omega_{1}}}{\Omega_{3}^{2}}.
\end{equation}

Using the EPR parameters [Eq. (25)] and the ground-state wave function of the superradiant phase [Eq. (45)], we can obtain the the two EPR steering parameters
\begin{equation}
E_{13}=\frac{\sin2\theta(\Omega_{1}-\Omega_{3})[\sin2\theta(\Omega_{1}-\Omega_{3}) -4\Omega_{1}\Omega_{3}]+4\Omega_{1}\Omega_{3}}{4\Omega_{1}\Omega_{3}[(\Omega_{1}+\Omega_{3})
+\cos2\theta(\Omega_{1}-\Omega_{3})]},
\end{equation}
\begin{equation}
E_{31}=\frac{\sin2\theta(\Omega_{1}-\Omega_{3})[\sin2\theta(\Omega_{1}-\Omega_{3}) -4\Omega_{1}\Omega_{3}]+4\Omega_{1}\Omega_{3}}{4\Omega_{1}\Omega_{3}[(\Omega_{1}+\Omega_{3})
-\cos2\theta(\Omega_{1}-\Omega_{3})]},
\end{equation}
where the $\sin2\theta$ and $\cos2\theta$ can be obtained by the Eq. (39).

Considering the cases of $\omega^2_{1}> \omega^2_{3}+4\eta\omega_3$ and $\tan2\theta<0$, we have
\begin{align}
\sin2\theta=\pm\sqrt{\frac{16\zeta^{2}\omega_{1}\omega_{3}}{(\omega_{1}^{2}-4\eta\omega_3-\omega_{3}^{2})^{2}+16\zeta^{2}\omega_{1}\omega_{3}}}, \\
\cos2\theta=\mp\sqrt{\frac{(\omega_{3}^{2}+4\eta\omega_3-\omega_{1}^{2})^{2}}{(\omega_{1}^{2}-4\eta\omega_3-\omega_{3}^{2})^{2}+16\zeta^{2}\omega_{1}\omega_{3}}}.
\end{align}

%%%%%%%%%%%%%%%%%%%%%%%%%%%%%%%%%%%%%%%%%%%%%%%%%%%%%%%%%%%%%%%%%%%%%%
%In what follows we consider the case of $\omega_{1}<\omega_{2}$. In this case $\tan2\theta>0$, we have
%\begin{align}
%\sin2\theta=\pm\sqrt{\frac{16\lambda^{2}\omega_{1}\omega_{2}}{(\omega_{1}^{2}-\omega_{2}^{2})^{2}+16\lambda^{2}\omega_{1}\omega_{2}}}\notag\\
%\cos2\theta=\pm\sqrt{\frac{(\omega_{2}^{2}-\omega_{1}^{2})^{2}}{(\omega_{1}^{2}-\omega_{2}^{2})^{2}+16\lambda^{2}\omega_{1}\omega_{2}}}
%\end{align}
%%%%%%%%%%%%%%%%%%%%%%%%%%%%%%%%%%%%%%%%%%%%%%%%%

\begin{figure}[t]
  \centerline{
  \includegraphics[width=0.5\textwidth]{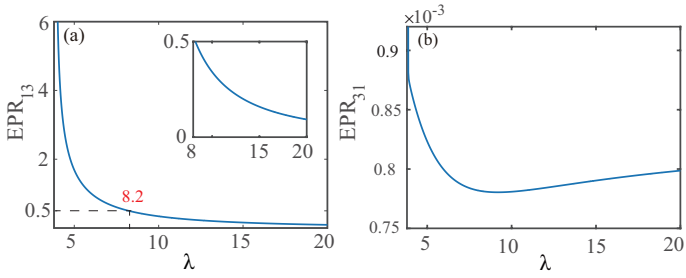}}
  \caption{EPR steering with the impurity atom versus the coupling strength $\lambda$ between the cavity field and the condensed atoms, when
the mixing angle obeys the condition $\sin 2\theta >0$. The numerical simulation parameters are $\omega=400\omega_{r},\ \chi=0.1\omega_{r},\  \chi^{"}=0.1\omega_{r},\ \kappa=0.5\omega_{r},\ \xi_{1}=0.001\omega_{r}$, and $\delta=0.5\omega_{r}$.} \label{fig5}
\end{figure}

In what follows we numerically analyze the EPR steering properties of the ground-state wave function of the superradiant phase in terms of the analytical expressions in Eqs. (30) and (31) for the case of  $\omega^2_{1}> \omega^2_{3}+4\eta\omega_3$ with $\sin 2\theta>0$ and $\cos 2\theta<0$, respectively.

In Fig. 5 we plot the two EPR parameters as the function of the coupling strength between the cavity field and condensed atoms in the presence of the impurity atom with the mixing angle obeys the condition  $\sin 2\theta >0$. Here we take  $\omega=400\omega_{r},\ \chi=0.1\omega_{r},\ \chi^{"}=0.1\omega_{r},\ \kappa=0.5\omega_{r},\ \xi_{1}=0.001\omega_{r}$, and $\delta=0.5\omega_{r}$.
From Figs. 5 (a) and (b) we can see that the two-way EPR steering occurs between the cavity field and condensed atoms in the regime of $\lambda> 8.2\omega_{r}$ while the one-way EPR steering can happen in the regime of $\lambda_c<\lambda< 8.2\omega_{r}$.

In Fig. 6 we plot the two EPR parameters as the function of the coupling strength between the cavity field and condensed atoms in the absence of the impurity atom, when the mixing angle obeys the condition  $\sin 2\theta >0$. Here we take  $\omega=400\omega_{r},\ \chi=0.1\omega_{r},$ and $\chi^{"}=0.1\omega_{r}$.
From Figs. 6 (a) and (b) we can see that the two-way EPR steering occurs between the cavity field and condensed atoms in the whole regime of $\lambda> \lambda_c$.

Similar to the case of the normal phase, we can compare the EPR steering results  of Fig. 5 with the impurity atom and Fig.6 without the impurity atom, we can find that the presence of the impurity atom can induce the transition from the two-way EPR steering between the cavity field and condensed atoms to the one-way EPR steering. Therefore, the impurity atom can also be regarded as a single atom switch, realizing the transition from the two-way EPR steering to the one-way EPR steering in the superradiant  phase. That is to say, when the impurity-BEC coupling is switched on, the one-way EPR steering from the cavity field to condensed atoms appears, whereas when the impurity-BEC coupling is switched off, the two-way EPR steering appears.

In Fig. 7 we plot the two EPR parameters as the function of the coupling strength between the cavity field and condensed atoms in the presence of the impurity atom, when the mixing angle obeys the condition  $\sin 2\theta >0$. Here we take  $\omega=400\omega_{r},\ \chi=0.1\omega_{r},\ \chi^{"}=0.1\omega_{r},\ \kappa=0.5\omega_{r},\ \xi_{1}=0.001\omega_{r}$, and $\delta=0.5\omega_{r}$.
From Figs. 7 (a) and (b) we can see that the two-way EPR steering occurs in the regime of $\lambda> 12.4\omega_{r}$ while the one-way EPR steering can happen  from condensed atoms to the cavity field in the regime of $\lambda_c<\lambda< 12.4\omega_{r}$.

\begin{figure}[t]
  \centerline{
  \includegraphics[width=0.5\textwidth]{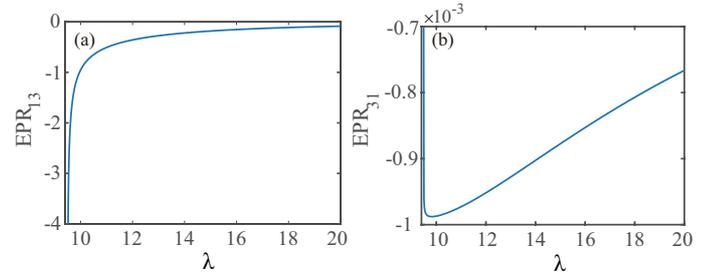}}
  \caption{EPR steering without the impurity atom versus  the coupling strength $\lambda$ between the cavity field and the condensed atoms, when
the mixing angle obeys the condition $\sin 2\theta >0$. The numerical simulation parameters are $\omega=400\omega_{r},\ \chi=0.1\omega_{r}$, and $\chi^{"}=0.1\omega_{r}$.} \label{fig5}
\end{figure}

In Fig. 8 we plot the two EPR parameters as the function of the coupling strength between the cavity field and condensed atoms in the absence of the impurity atom, when the mixing angle obeys the condition  $\sin 2\theta<0$. Here we take $\omega=400\omega_{r},\ \chi=0.1\omega_{r}$, and $\chi^{"}=0.1\omega_{r}$.
From Figs. 8 (a) and (b) we can see that the two-way EPR steering occurs in the regime of $\lambda> 17.4\omega_{r}$ while the one-way EPR steering  can happen  from  condensed atoms to the cavity field in the regime of $\lambda_c<\lambda< 17.4\omega_{r}$.

Comparing the EPR steering results of Fig. 7 with Fig. 5 in the region of the one-way EPR steering, it is straightforward to see that one can change the direction of the one-way EPR steering through adjusting the sign of the mixing angle function $\sin 2\theta$. When $\sin 2\theta>0$, the cavity field can unidirectionally steer the condensed atoms while when $\sin 2\theta<0$, the condensed atoms can unidirectionally steer the cavity field.

From the above numerical analysis we can conclude that in the superradiant phase, when the mixing angle obeys $\sin 2\theta >0$, there exist either one-way or two-way EPR steering in different regime of the impurity-BEC coupling strength. In the whole regime of the impurity-BEC coupling strength, there is one-way EPR steering from the cavity field to condensed atoms while in the regime of far away from  the critical point of the QPT, there is two-way EPR steering between the cavity field and condensed atoms.
When the mixing angle obeys $\sin 2\theta <0$, there also exist either one-way or two-way EPR steering in different regime of the impurity-BEC coupling strength.  In the whole regime of the impurity-BEC coupling strength, there is one-way EPR steering from condensed atoms to the cavity field while in the regime of far away from  the critical point of the QPT, there is two-way EPR steering between the cavity field and condensed atoms. The single impurity can also be regarded as a single-atom switch to control the EPR steering direction. When the mixing angle obeys $\sin 2\theta >0$, the two-way EPR steering can be turned to the one-way EPR steering through switching off the impurity-BEC coupling whli the one-way EPR steering can be changed to the two-way EPR steering through switching on the impurity-BEC coupling.

\begin{figure}[t]
  \centerline{
  \includegraphics[width=0.5\textwidth]{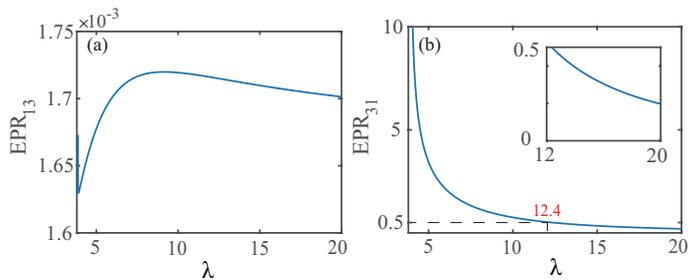}}
  \caption{EPR steering with the impurity atom versus  the coupling strength $\lambda$ between the cavity field and the condensed atoms, when
the mixing angle obeys the condition $\sin 2\theta <0$. The numerical simulation parameters are $\omega=400\omega_{r},\ \chi=0.1\omega_{r} ,\ \chi^{"}=0.1\omega_{r},\ \kappa=0.5\omega_{r},\ \xi_{1}=0.001\omega_{r}$, and $\delta=0.5\omega_{r}$.} \label{fig5}
\end{figure}

\begin{figure}[t]
  \centerline{
  \includegraphics[width=0.5\textwidth]{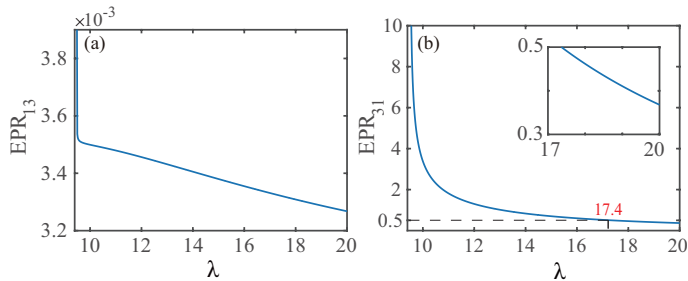}}
  \caption{EPR steering without the impurity atom versus  the coupling strength $\lambda$ between the cavity field and the condensed atoms, when
the mixing angle obeys the condition $\sin 2\theta <0$. The numerical simulation parameters are $\omega=400\omega_{r},\ \chi=0.1\omega_{r}$, and $\chi^{"}=0.1\omega_{r}$.} \label{fig5}
\end{figure}

\section{ Witness QPT by EPR steering }

In this section, we numerically investigate the relationship between the EPR steering and the QPT in the GDM generated from the impurity-doped cavity-BEC system. We show that the two EPR steering parameters can be used as witnessing parameters of the  QPT through a sudden change at the critical point of the QPT in the GDM.

In Fig. 9 we  plot the two EPR parameters as the function of the coupling strength between the cavity field and condensed atoms in the presence of the impurity atom,  when the mixing angle obeys the condition  $\sin 2\theta<0$ and  $\sin 2\theta<0$, respectively. Here we take  $\omega=400\omega_{r},\ \chi=0.1\omega_{r},\  \chi^{"}=0.1\omega_{r},\ \kappa=0.5\omega_{r},\ \xi_{1}=0.001\omega_{r}$, $\delta=0.5\omega_{r}$, $\sin 2\theta >0$,  and $\sin 2\theta <0$.
Figs. 9 (a) and (b)  correspond to the case  of $\sin 2\theta>0$  while Figs. 9 (c) and (d)  correspond to the case  of $\sin 2\theta<0$.
From Fig. 9 we can see that both of  the two EPR steering parameters  $E_{12}$ and  $E_{21}$  have a sudden change at the critical point of the QPT $\lambda=\lambda_c$ with $\lambda_c=3.87\omega_{r}$.
Hence such a sudden change of the two EPR steering parameters can witness happening of the QPT induced by the coupling between the cavity field and condensed atoms.
From Figs. 9 (a) and (b), we can find that the EPR steering parameter  $E_{12}$ is  more sensitive to the QPT than $E_{21}$  due to $E_{12} \gg E_{21}$ for $\sin 2\theta>0$. Hence $E_{12}$ is a better witnessing parameter with respect to $E_{21}$ for the case of  $\sin 2\theta>0$. For $\sin 2\theta<0$, the EPR steering parameter  $E_{21}$ is  more sensitive to the QPT than  $E_{12}$ due to $E_{21} \gg E_{12}$, shown in Figs. 9 (c) and (c). Hence $E_{21}$ is  a better witnessing parameter with respect to $E_{12}$  for the case of  $\sin 2\theta<0$.

\begin{figure}[t]
  \centerline{
  \includegraphics[width=0.5\textwidth]{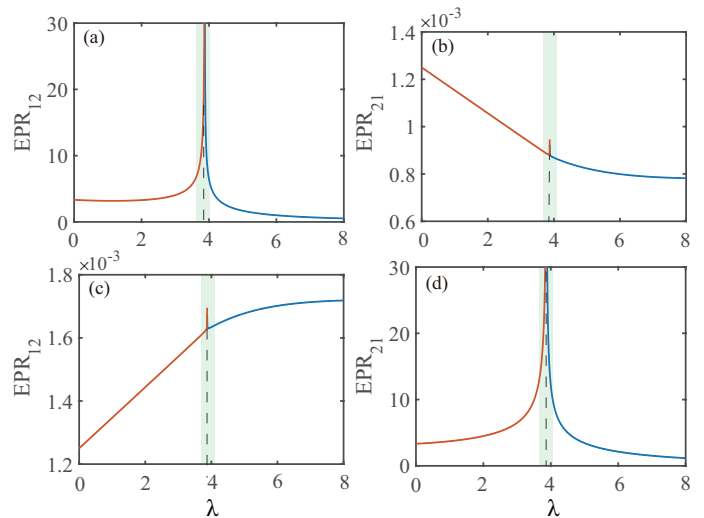}}
  \caption{Witnessing QPT by EPR steering parameters in the presence of the impurity atom. EPR steering with the impurity atom versus the coupling strength $\lambda$ between the cavity field and the condensed atoms with various mixing angle condition: (a,b) $\sin 2\theta >0$, (c,d) $\sin 2\theta <0$. The numerical simulation parameters are $\omega=400\omega_{r},\ \chi=0.1\omega_{r},\ \chi^{"}=0.1\omega_{r},\ \kappa=0.5\omega_{r},\ \xi_{1}=0.001\omega_{r}$, and $\delta=0.5\omega_{r}$.} \label{fig5}
\end{figure}

In Fig. 10 we  plot  the two EPR parameters as the function of the coupling strength between the cavity field and condensed atoms in the absence of the impurity atom,  when the mixing angle obeys the condition  $\sin 2\theta<0$ and $\sin 2\theta<0$. Here we take  $\omega=400\omega_{r},\ \chi=0.1\omega_{r},\ \chi^{"}=0.1\omega_{r}$, $\sin 2\theta >0$, and $\sin 2\theta <0$, respectively.
From Fig. 10, we can find the EPR steering parameter  $E_{12}$ is  a better witnessing parameter with respect to $E_{21}$ for $\sin 2\theta>0$ shown in Figs. 10 (a) and (b), and the EPR steering parameter  $E_{21}$ is  a better witnessing parameter with respect to $E_{12}$  for the case of  $\sin 2\theta<0$ shown in Figs. 10 (c) and (d). In addition, we can find that the impurity atom can decrease the critical coupling strength of the QPT by comparing the results in Fig. 9 and Fig. 10.

\begin{figure}[t]
  \centerline{
  \includegraphics[width=0.5\textwidth]{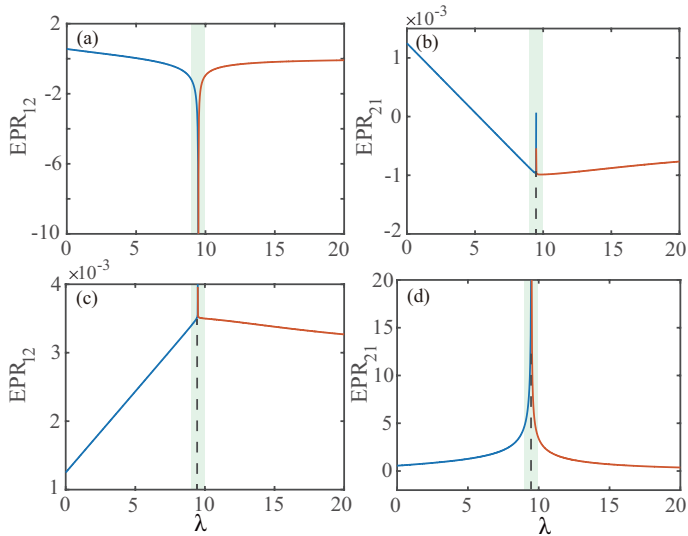}}
  \caption{Witnessing QPT by EPR steering parameters in the absence of the impurity atom. EPR steering with the impurity atom versus the coupling strength $\lambda$ between the cavity field and the condensed atoms with various mixing angle condition: (a,b) $\sin 2\theta >0$, (c,d) $\sin 2\theta <0$. The numerical simulation parameters are $\omega=400\omega_{r},\ \chi=0.1\omega_{r}$, and $\chi^{"}=0.1\omega_{r}$.} \label{fig5}
\end{figure}

\section{Conclusions}

In this work, we have studied EPR steering properties of the normal and superradiant phases in the GDM generated by a cavity BEC doped with a single impurity. We have found that the ground states of the normal and  superradiant phases of the GDM exhibit much rich EPR steerability. In the normal phase, there exist either one-way EPR steering from the cavity field to condensed atoms or one-way EPR steering from condensed atoms to the cavity field, and the direction of the EPR steering can be controlled through adjusting the mixing angle. However, in the normal phase there is not two-way EPR steering between the cavity field and condensed atoms. In the superradiant phase, the EPR steering is more complex than that in the normal phase. There exist either one-way or two-way EPR steering in different regime of the impurity-BEC coupling strength. We have shown that the single impurity can be regarded as a single-atom switch to control the EPR steering direction. It can be used either to control  the steering direction of one-way EPR steering or to manipulate the transition between one-way and two-way through switching on or off the impurity-BEC coupling.
We have also proved that the EPR steering parameters can witness the QPT occurrence through the sudden change of the EPR steering parameters at the critical point of the QPT.  The EPR steering between the flying photons and localized condensed atoms, is more useful for the practical quantum communication network. Our results open a new way to manipulate macroscopic EPR steering and  witness the QPT  between the normal phase and the superradiant phase in the GDM through a microscopic impurity atom.

\begin{acknowledgements}
H.J. was supported by the  the National Natural Science Foundation of China  (NSFC) (Grants Nos. 11935006 and   11774086) and the Science and Technology Innovation Program of Hunan Province (Grant No. 2020RC4047). L.-M.K. was supported by NSFC (Grant Nos. 1217050862 and 11775075). Y.-F.J. was supported by the NSFC (Grant No. 12147156), the China Postdoctoral Science Foundation (Grant No. 2021M701176) and the Science and Technology Innovation Program of Hunan Province (Grant No. 2021RC2078). B.J.L. was supported by Postgraduate Scientific Research Innovation Project of Hunan Province (Grant No. CX20210471).
\end{acknowledgements}

%
%\appendix
%\begin{widetext}

%\section{A}
%
%\end{widetext}


\begin{thebibliography}{99}
\bibitem{1} H.~M.~Wiseman, S.~J.~Jones, and A.~C.~Doherty, Steering, Entanglement, Nonlocality, and the Einstein-Podolsky-Rosen Paradox, Phys. Rev. Lett. \textbf{98}, 140402 (2007). https://doi.org/10.1103/PhysRevLett.98.140402

\bibitem{2} E.~Schr\"{o}dinger, Discussion of Probability Relations between Separated Systems, Math. Proc. Camb. Philos. Soc. \textbf{31}, 555--563 (1935).
    https://doi.org/10.1017/S0305004100013554

\bibitem{3} E.~Schr\"{o}dinger, Probability relations between separated systems, Math. Proc. Camb. Philos. Soc. \textbf{32}, 446--452 (1936).
    https://doi.org/10.1017/S0305004100019137

\bibitem{4} A.~Einstein, B.~Podolsky, and N.~Rosen, Can Quantum-Mechanical Description of Physical Reality Be Considered Complete? Phys. Rev. \textbf{47}, 777--780 {1935}.
    https://doi.org/10.1103/PhysRev.47.777

\bibitem{5} R.~Uola, A.~C.~S.~Costa, H.~C.~Nguyen, and O.~G\"{u}hne, Quantum steering, Rev. Mod. Phys. \textbf{92}, 015001 (2020).
    https://doi.org/10.1103/RevModPhys.92.015001

\bibitem{6} M.~D.~Reid, P.~D.~Drummond, W.~P.~Bowen, E.~G.~Cavalcanti, P.~K.~Lam, H.~A.~Bachor, U.~L.~Andersen, and G.~Leuchs, ColloColloquium: The Einstein-Podolsky-Rosen paradox: From concepts to applications, Rev. Mod. Phys. \textbf{81}, 1727 (2009).
    https://doi.org/10.1103/RevModPhys.81.1727

\bibitem{7} C.~Branciard, E.~G.~Cavalcanti, S.~P.~Walborn, V.~Scarani, and H.~M.~Wiseman, One-sided device-independent quantum key distribution: Security, feasibility, and the connection with steering, Phys. Rev. A \textbf{85}, 010301(R) (2012).
    https://doi.org/10.1103/PhysRevA.85.010301

\bibitem{8} M.~Hillery, V. Bu\v{z}ek, and A. Berthiaume, Quantum secret sharing,
Phys. Rev. A \textbf{59}, 1829 (1999).
https://doi.org/10.1103/PhysRevA.59.1829


\bibitem{10} S.~Gaertner, C.~Kurtsiefer, M.~Bourennane, and H.~Weinfurter, Experimental Demonstration of Four-Party Quantum Secret Sharing, Phys. Rev. Lett. \textbf{98}, 020503 (2007).
    https://doi.org/10.1103/PhysRevLett.98.020503

\bibitem{11} A.~M.~Lance, T.~Symul, W.~P.~Bowen, T.~Tyc, B.~C.~Sanders, and P.~K.~Lam, Continuous variable (2, 3) threshold quantum secret sharing schemes, New J. Phys. \textbf{5}, 4 (2003).
    https://doi.org/10.1088/1367-2630/5/1/304

\bibitem{12} Y.-A.~Chen, A.-N.~Zhang, Z.~Zhao, X.-Q.~Zhou, C.-Y.~Lu, C.-Z.~Peng, T.~Yang, and J.-W.~Pan, Experimental Quantum Secret Sharing and Third-Man Quantum Cryptography, Phys. Rev. Lett. \textbf{95}, 200502 (2005).
    https://doi.org/10.1103/PhysRevLett.95.200502

\bibitem{13} Y.~Xiang, I.~Kogias, G.~Adesso, and Q.~Y.~He, Multipartite Gaussian steering: Monogamy constraints and quantum cryptography application, Phys. Rev. A \textbf{95}, 010101(R) (2017).
    https://doi.org/10.1103/PhysRevA.95.010101

\bibitem{14} D.~Cavalcanti, P.~Skrzypczyk, G.~H.~Aguilar, R.~V.~Nery, P.~H.~S.~Ribeiro, and S.~P.Walborn, Detection of entanglement in asymmetric quantum networks and multipartite quantum steering, Nat. Commun. \textbf{6}, 7941 (2015).
    https://doi.org/10.1038/ncomms8941

%%%%%%%%%%%%%%%
\bibitem{15} S.~L.~W.~Midgley, A.~J.~Ferris, and M.~K.~Olsen, Asymmetric Gaussian steering: When Alice and Bob disagree, Phys. Rev. A \textbf{81}, 022101 (2010).
    https://doi.org/10.1103/PhysRevA.81.022101

\bibitem{16} M.~K.~Olsen and J.~F.~Corney, Non-Gaussian continuousvariable entanglement an steering, Phys. Rev. A \textbf{87}, 033839 (2013).
    https://doi.org/10.1103/PhysRevA.87.033839

\bibitem{17} Q.~Y.~He and Z.~Ficek, Einstein-Podolsky-Rosen paradox and quantum steering in a three-mode optomechanical system, Phys. Rev. A \textbf{89}, 022332 (2014).
    https://doi.org/10.1103/PhysRevA.89.022332

\bibitem{18} E.~G.~Cavalcanti, Q.~Y.~He, M.~D.~Reid, and H.~M.~Wiseman, Unified criteria for multipartite quantum nonlocality, Phys. Rev. A \textbf{84}, 032115 (2011).
    https://doi.org/10.1103/PhysRevA.84.032115

\bibitem{19} Q.~Y.~He and M.~D.~Reid, Genuine Multipartite Einstein-Podolsky-Rosen Steering, Phys. Rev. Lett. \textbf{111}, 250403 (2013).
    https://doi.org/10.1103/PhysRevLett.111.250403

\bibitem{20} Q.~Y.~He, L.~Rosales-Z\'{a}rate, G.~Adesso, and M.~D.~Reid, Secure Continuous Variable Teleportation and Einstein-Podolsky-Rosen Steering, Phys. Rev. Lett. \textbf{115}, 180502 (2015).
    https://doi.org/10.1103/PhysRevLett.115.180502

\bibitem{21} I.~Kogias, A.~R.~Lee, S.~Ragy, and G.~Adesso, Quantification of Gaussian Quantum Steering, Phys. Rev. Lett. \textbf{114}, 060403 (2015).
    https://doi.org/10.1103/PhysRevLett.114.060403

\bibitem{22} V.~H\"{a}ndchen, T.~Eberle, S.~Steinlechner, A.~Samblowski, T.~Franz, R.~F.~Werner, and R.~Schnabel, Observation of one-way Einstein-Podolsky-Rosen steering, Nat. Photon. \textbf{6}, 596--599 (2012).
    https://doi.org/10.1038/nphoton.2012.202

\bibitem{23} S.~Armstrong, M.~Wang, R.~Y.~Teh, Q.~H.~Gong, Q.~Y.~He, J.~Janousek, H.-A.~Bachor, M.~D.~Reid, and P.~K.~Lam, Multipartite Einstein-Podolsky-Rosen steering and genuine tripartite entanglement with optical networks, Nat. Phys. \textbf{11}, 167--172 (2015). %%same as rer[46]
    https://doi.org/10.1038/nphys3202

\bibitem{24} M.~D.~Reid, Demonstration of the Einstein-Podolsky-Rosen paradox using nondegenerate parametric amplification, Phys. Rev. A \textbf{40}, 913 (1989).
    https://doi.org/10.1103/PhysRevA.40.913

\bibitem{25} M.~K.~Olsen, Controlled Asymmetry of Einstein-Podolsky-Rosen Steering with an Injected Nondegenerate Optical Parametric Oscillator, Phys. Rev. Lett. \textbf{119}, 160501 (2017).
    https://doi.org/10.1103/PhysRevLett.119.160501

\bibitem{26} M.~K.~Olsen, Entanglement and asymmetric steering over two octaves of frequency difference, Phys. Rev. A \textbf{96}, 063839 (2017).
    https://doi.org/10.1103/PhysRevA.96.063839

\bibitem{27} M.~K.~Olsen, Third-harmonic entanglement and Einstein-Podolsky-Rosen steering over a frequency range of more than an octave, Phys. Rev. A \textbf{97}, 033820 (2018).
    https://doi.org/10.1103/PhysRevA.97.033820

\bibitem{ChenAX} Y.~Liu, S.~L.~Liang, G.~R.~Jin, Y.~B.~Yu, and A.~X.~Chen, Einstein-Podolsky-Rosen steering in spontaneous parametric down-conversion cascaded with a sum-frequency generation, Phys. Rev. A \textbf{102}, 052214 (2020).
    https://doi.org/10.1103/PhysRevA.102.052214

\bibitem{TehRT} R.~Y.~Teh, M.~Gessner, M.~D.~Reid, and M.~Fadel, Full multipartite steering inseparability, genuine multipartite steering, and monogamy for continuous-variable systems, Phys. Rev. A \textbf{105}, 012202 (2022).
    https://doi.org/10.1103/PhysRevA.105.012202

\bibitem{Gianani} I.~Gianani, V.~Berardi, and M.~Barbieri, Witnessing quantum steering by means of the Fisher information, Phys. Rev. A \textbf{105}, 022421 (2022).
    https://doi.org/10.1103/PhysRevA.105.022421

\bibitem{DaiTZ} T.-Z.~Dai, Y.~Fan, and L.~Qiu, Complementary relation between tripartite entanglement and the maximum steering inequality violation, Phys. Rev. A \textbf{105}, 022425 (2022).
    https://doi.org/10.1103/PhysRevA.105.022425

\bibitem{ZhuJ} J.~Zhu, M.-J.~Hu, C.-F.~Li, G.-C.~Guo, and Y.-S.~Zhang, Einstein-Podolsky-Rosen steering in two-sided sequential measurements with one entangled pair, Phys. Rev. A \textbf{105}, 032211 (2022).
    https://doi.org/10.1103/PhysRevA.105.032211

\bibitem{ZhangJ} J.~Zhang, K.~He, Y.~Zhang, Y.-Y.~Hao, J.-C.~Hou, F.-P.~Lan, and B.-N.~Niu, Detecting the steerability bounds of generalized Werner states via a backpropagation neural network, Phys. Rev. A \textbf{105}, 032408 (2022).
    https://doi.org/10.1103/PhysRevA.105.032408

\bibitem{DesignolleS} S.~Designolle, Robust genuine high-dimensional steering with many measurements, Phys. Rev. A \textbf{105}, 032430 (2022). https://doi.org/10.1103/PhysRevA.105.032430;
S. Designolle, V. Srivastav, R. Uola, N. H. Valencia, W. McCutcheon, M. Malik, and N. Brunner, Phys. Rev. Lett. \textbf{126}, 200404 (2021). https://doi.org/10.1103/PhysRevLett.126.200404

\bibitem{SeifertLM} L.~M.~Seifert, K.~Beyer, K.~Luoma, and W.~T.~Strunz, Quantum steering on IBM quantum processors, Phys. Rev. A \textbf{105}, 042413 (2022).
    https://doi.org/10.1103/PhysRevA.105.042413

%%%%%%%%%%%%%%%%

\bibitem{ZhengSS}  S.-S.~Zheng, F.-X.~Sun, H.-Y.~Yuan, Z.~Ficek, Q.-H.~Gong, and Q.-Y.~He, Enhanced entanglement and asymmetric EPR steering between magnons, Sci. China Phys. Mech. Astron. \textbf{64}, 210311 (2021).
    https://doi.org/10.1007/s11433-020-1587-5

\bibitem{KongD} D.~Kong, J.~Xu, Y.~Tian, F.~Wang, and X.~M.~Hu, Remote asymmetric Einstein-Podolsky-Rosen steering of magnons via a single pathway of Bogoliubov dissipation, Phys. Rev. Research \textbf{4}, 013084 (2022).
    https://doi.org/10.1103/PhysRevResearch.4.013084

\bibitem{HanX} X.~H.~Han, Y.~Xiao, H.~C.~Qu, R.~H.~He, X.~Fan, T.~Qian, and Y.~J.~Gu, Sharing quantum steering among multiple Alices and Bobs via a two-qubit Werner state, Quantum Inf Process \textbf{20}, 278 (2021).
    https://doi.org/10.1007/s11128-021-03211-z

\bibitem{ZengQ} Q.~Zeng, J.~W.~Shang, H.~C.~Nguyen, and X.~D.~Zhang, Reliable experimental certification of one-way Einstein-Podolsky-Rosen steering, Phys. Rev. Research \textbf{4}, 013151 (2022).
    https://doi.org/10.1103/PhysRevResearch.4.013151

\bibitem{ZhaoSQ} S.~Q.~Zhao, Y.~M.~Long, M.~X.~Zhang, T.~Y.~Zheng, and X.~Zhang, Genuine tripartite entanglement in the dynamical Casimir coupled waveguides, Quantum Inf Process \textbf{20}, 308 (2021).
    https://doi.org/10.1007/s11128-021-03247-1

\bibitem{TanHT} H.~T.~Tan and J.~Li, Einstein-Podolsky-Rosen entanglement and asymmetric steering between distant macroscopic mechanical and magnonic systems, Phys. Rev. Research \textbf{3}, 013192 (2021).
    https://doi.org/10.1103/PhysRevResearch.3.013192

\bibitem{ChengWW} W.~W.~Cheng, B.~W.~Wang, L.~Y.~Gong, and S.~M.~Zhao, Dynamics of Einstein-Podolsky-Rosen steering in Heisenberg model under decoherence, Quantum Inf Process \textbf{20}, 371 (2021).
    https://doi.org/10.1007/s11128-021-03309-4

\bibitem{PanGZ} G.-Z.~Pan, M.~Yang, H.~Yuan, G.~Zhang, and J.-L.~Zhao, Nonlinear steering criteria for arbitrary two-qubit quantum systems, Quantum Inf Process \textbf{20}, 48 (2021).
    https://doi.org/10.1007/s11128-020-02954-5

%%%%%%%%%%%%%%%%%%%%

\bibitem{28} B.~Wittmann, S.~Ramelow, F.~Steinlechner, N.~K.~Langford, N.~Brunner, H.~M.~Wiseman, R.~Ursin, and A.~Zeilinger, Loophole-free Einstein-Podolsky-Rosen experiment via quantum steering, New J. Phys. \textbf{14}, 053030 (2012).
    https://doi.org/10.1088/1367-2630/14/5/053030

\bibitem{29} D.~J.~Saunders, S.~J.~Jones, H.~M.~Wiseman, and G.~J.~Pryde, Experimental EPR-steering using Bell-local states, Nat. Phys. \textbf{6}, 845--849 (2010).
    https://doi.org/10.1038/nphys1766

%\bibitem{30} S.~Armstrong, M.~Wang, R.~Y.~Teh, Q.~H.~Gong, Q.~Y.~He, J.~Janousek, H.-A.~Bachor, M.~D.~Reid, P.~K.~Lam, Multipartite Einstein-Podolsky-Rosen steering and genuine tripartite entanglement %with optical networks, Nat. Phys. \textbf{11}, 167--172 (2015).    https://doi.org/10.1038/nphys3202

\bibitem{31} T.~Wasak, J.~Chwede\'{n}czuk, Bell Inequality, Einstein-Podolsky-Rosen Steering, and Quantum Metrology with Spinor Bose-Einstein Condensates, Phys. Rev. Lett. \textbf{120}, 140406 (2018).
    https://doi.org/10.1103/PhysRevLett.120.140406

\bibitem{32} N.~Tischler, F.~Ghafari, T.~J.~Baker, S.~Slussarenko, R.~B.~Patel, M.~M.~Weston, S.~Wollmann, L.~K.~Shalm, V.~B.~Verma, S.~W.~Nam, H.~C.~Nguyen, H.~M.~Wiseman, and G.~J.~Pryde, Conclusive Experimental Demonstration of One-Way Einstein-Podolsky-Rosen Steering, Phys. Rev. Lett. \textbf{121}, 100401 (2018).
    https://doi.org/10.1103/PhysRevLett.121.100401

\bibitem{33} P.~Kunkel, M.~Pr\"{u}fer, H.~Strobel, D.~Linnemann, A.~Fr\"{o}lian, T.~Gasenzer, M.~G\"{a}rttner, and M.~K.~Oberthaler, Spatially distributed multipartite entanglement enables EPR steering of atomic clouds, Science \textbf{360}, 413--416 (2018).
    https://doi.org/10.1126/science.aao2254

\bibitem{34} K.~Lange, J.~Peise, B.~L\"{u}cke, I.~Kruse, G.~Vitagliano, I.~Apellaniz, M.~Kleinmann, G.~T\'{o}th, and C.~Klempt, Entanglement between two spatially separated atomic modes, Science \textbf{360}, 416--418 (2018).
    https://doi.org/10.1126/science.aao2035

\bibitem{35} M.~Fadel, T.~Zibold, B.~D\'{e}camps, and P.~Treutlein, Spatial entanglement patterns and Einstein-Podolsky-Rosen steering in Bose-Einstein condensates, Science \textbf{360}, 409--413 (2018).
    https://doi.org/10.1126/science.aao1850

%%%%%%%%%%%%
\bibitem{36} X.~Y.~Huang, E.~Zeuthen, Q.~H.~Gong, and Q.~Y.~He, Engineering asymmetric steady-state Einstein-Podolsky-Rosen steering in macroscopic hybrid systems, Phys. Rev. A \textbf{100}, 012318 (2019).
    https://doi.org/10.1103/PhysRevA.100.012318

\bibitem{37} L.~Zeng, R.~Ma, H.~Wen, M.~H.~Wang, J.~Liu, Z.~Z.~Qin, and X.~L.~Su, Deterministic distribution of orbital angular momentum multiplexed continuous-variable entanglement and quantum steering, Photonics Research  \textbf{10}, 777--785 (2022).
    https://doi.org/10.1364/PRJ.442925

\bibitem{38} Z.-Y.~Hao, K.~Sun, Y.~Wang, Z.-H.~Liu, M.~Yang, J.-S.~Xu, C.-F.~Li, and G.-C.~Guo, Demonstrating Shareability of Multipartite Einstein-Podolsky-Rosen Steering, Phys. Rev. Lett. \textbf{128}, 120404 (2021).
    https://doi.org/10.1103/PhysRevLett.128.120402

\bibitem{LiuSL}  S.~H.~Liu, D.~M.~Han, N.~Wang, Y.~Xiang, F.~X.~Sun, M.~H.~Wang, Z.~Z.~Qin, Q.~H.~Gong, X.~L.~Su, and Q.~Y.~He, Experimental Demonstration of Remotely Creating Wigner Negativity via Quantum Steering, Phys. Rev. Lett. \textbf{128}, 200401 (2022).
    https://doi.org/10.1103/PhysRevLett.128.200401

%%%%%%%%%%%%%%%%

\bibitem{Brennecke2007}  F.~Brennecke, T.~Donner, S.~Ritter, T.~Bourdel,  M.~K\"{o}hl, and T.~Esslinger, Cavity QED with a Bose-Einstein  condensate, Nature (London) \textbf{450}, 268--271 (2007).
    https://doi.org/10.1038/nature06120

\bibitem{Colombe2007}  Y.~Colombe, T.~Steinmetz, G.~Dubois, F.~Linke, D.~Hunger, and J.~Reichel, Strong atom-field coupling for Bose-Einstein condensates in an optical cavity on a chip, Nature (London) \textbf{450}, 272--276 (2007).
    https://doi.org/10.1038/nature06331

\bibitem{Dimer} F.~Dimer, B.~Estienne, A.~S.~Parkins, and H.~J.~Carmichael, Proposed realization of the Dicke-model quantum phase transition in an optical cavity QED system, Phys. Rev. A \textbf{75}, 013804 (2007).
    https://doi.org/10.1103/PhysRevA.75.013804

\bibitem{Baumann1}  K.~Baumann, C.~Guerlin, F.~Brennecke, and T.~Esslinger, Dicke quantum phase transition with a superfluid gas in an optical cavity, Nature (London) \textbf{464}, 1301 (2010).
    https://doi.org/10.1038/nature09009

\bibitem{Yuan2}   J. B. Yuan, W. J Lu, Y. J. Song, and L. M. Kuang, Single-impurity-induced Dicke quantum phase transition in a cavity-Bose-Einstein condensate, Scientific Reports   \textbf{7}, 7404 (2018). https://doi.org/10.1038/s41598-017-07899-x

\bibitem{Keeling} J.~Keeling,  M.~J.~Bhaseen, and  B.~D.~Simons, Collective Dynamics of Bose-Einstein Condensates in Optical Cavities, Phys. Rev. Lett. \textbf{105}, 043001 (2010).
    https://doi.org/10.1103/PhysRevLett.105.043001

\bibitem{Nagy1}  D.~Nagy, G.~K\'{o}nya, G.~Szirmai, and P.~Domokos, Dicke-Model Phase Transition in the Quantum Motion of a Bose-Einstein Condensate in an Optical Cavity, Phys. Rev. Lett. \textbf{104}, 130401 (2010).
    https://doi.org/10.1103/PhysRevLett.104.130401

\bibitem{Nagy2} D.~Nagy, G.~Szirmai, and  P.~Domokos, Critical exponent of a quantum-noise-driven phase transition: The open-system Dicke model, Phys. Rev. A \textbf{84}, 043637 (2011).
    https://doi.org/10.1103/PhysRevA.84.043637

\bibitem{Baumann2}  K.~Baumann, R.~Mottl, F.~Brennecke, and  T.~Esslinger, Exploring Symmetry Breaking at the Dicke Quantum Phase Transition, Phys. Rev. Lett. \textbf{107}, 140402 (2011).
    https://doi.org/10.1103/PhysRevLett.107.140402

\bibitem{Bastidas}  V.~M.~Bastidas, C.~Emary, B.~Regler, and T.~Brandes, Nonequilibrium Quantum Phase Transitions in the Dicke Model, Phys. Rev. Lett. \textbf{108}, 043003 (2012).
    https://doi.org/10.1103/PhysRevLett.108.043003

\bibitem{Bhaseen} M.~J.~Bhaseen, J.~Mayoh, B.~D.~Simons, and  J.~Keeling, Dynamics of nonequilibrium Dicke models, Phys. Rev. A \textbf{85}, 013817 (2012).
    https://doi.org/10.1103/PhysRevA.85.013817

%%%%%%%%%%%%
\bibitem{Klinder} J. Klinder, H. Ke{\ss}ler, M. Wolke, L. Mathey, and A. Hemmerich, Dynamical phase transition in the open Dicke model, PNAS \textbf{112}, 3290 (2015).
    https://doi.org/10.1073/pnas.1417132112


\bibitem{Liu1} N.~Liu, J.~L.~Lian, J.~Ma, L.~T.~Xiao, G.~Chen, J.-Q.~Liang, S.~T.~Jia, Light-shift-induced quantum phase transitions of a Bose-Einstein condensate in an optical cavity, Phys. Rev. A \textbf{83}, 033601 (2011).
    https://doi.org/10.1103/PhysRevA.83.033601

\bibitem{Yuan}  J.~B.~Yuan, and  L.-M.~Kuang, Quantum-discord amplification induced by a quantum phase transition via a cavity Bose-Einstein-condensate system, Phys. Rev. A \textbf{87}, 024101 (2013).
    https://doi.org/10.1103/PhysRevA.87.024101

%%%%%%%%%%%%%%%
\bibitem{Ng} H.~T.~Ng and S.~Bose, Single-atom-aided probe of the decoherence of a Bose-Einstein condensate, Phys. Rev. A \textbf{78}, 023610 (2008).
    https://doi.org/10.1103/PhysRevA.78.023610

\bibitem{Balewski} J.~B.~Balewski, A.~T.~Krupp, A.~Gaj, D.~Peter, H.~P.~B\"{u}chler, R.~L\"{o}w, S.~Hofferberth, and T.~Pfau, Coupling a single electron to a Bose-Einstein condensate, Nature (London) \textbf{502}, 664--667 (2013).
    https://doi.org/10.1038/nature12592

\bibitem{Schmidt}  R.~Schmidt, H.~R.~Sadeghpour, and E.~Demler, Mesoscopic Rydberg impurity in an atomic quantum gas, Phys. Rev. Lett. \textbf{116}, 105302 (2016). https://doi.org/10.1103/PhysRevLett.116.105302

\bibitem{Johnson} T.~H.~Johnson, Y.~Yuan, W.~Bao, S.~R.~Clark, C.~Foot, and D.~Jaksch, Hubbard model for atomic impurities bound by the vortex lattice of a rotating Bose-Einstein condensate, Phys. Rev. Lett. \textbf{116}, 240402 (2016).
    https://doi.org/10.1103/PhysRevLett.116.240402

\bibitem{LiQIP}  Z.~Li,  L.-M.~Kuang, Controlling quantum coherence of a two-component Bose-Einstein condensate via an impurity atom, Quantum Inf. Process  \textbf{19}, 188 (2020). https://doi.org/10.1007/s11128-020-02689-3

\bibitem{YuanJB} J.-B.~Yuan, W.-J.~Lu, Y.-J.~Song, and L.-M.~Kuang, Single-impurity-induced Dicke quantum phase transition in a cavity-Bose-Einstein condensate, Sci. Rep. \textbf{7}, 7404 (2017).
    https://doi.org/10.1038/s41598-017-07899-x


\bibitem{SongYJ} Y.-J.~Song and L.-M.~Kuang, Controlling Decoherence Speed Limit of a Single Impurity Atom in a Bose-Einstein-Condensate Reservoir, Ann. Phys. (Berlin) \textbf{531}, 1800423 (2019).
    https://doi.org/10.1002/andp.201800423

\bibitem{Rammohan1}  S.~Rammohan, S.~Tiwari, A.~Mishra, A.~Pendse, A.~K.~Chauhan, R.~Nath, A.~Eisfeld, and S.~W\"{u}ster, Imaging the interface of a qubit and its quantum-many-body environment. Phys. Rev. A \textbf{104}, L060202 (2021). https://doi.org/10.1103/PhysRevA.104.L060202

\bibitem{Rammohan2}  S.~Rammohan, A.~K.~Chauhan, R.~Nath, A.~Eisfeld, and  S.~W\"{u}ster, Tailoring Bose-Einstein-condensate environments for a Rydberg impurity, Phys. Rev. A \textbf{103}, 063307 (2021).
    https://doi.org/10.1103/PhysRevA.103.063307

\bibitem{Schleier-Smith} M.~Schleier-Smith, Editorial: Hybridizing Quantum Physics and Engineering, Phys. Rev. Lett. \textbf{117}, 100001 (2016).
    https://doi.org/10.1103/PhysRevLett.117.100001

\bibitem{Kurizkia} G.~Kurizkia, P.~Bertetb, Y.~Kubob, K. M{\o}lmer, D.~Petrosyan, P.~Rabl, and J.~Schmiedmayer, Quantum technologies with hybrid systems, PNAS \textbf{112}, 3866 (2015).
    https://doi.org/10.1073/pnas.1419326112

\bibitem{Quirion} D.~Lachance-Quirion, Y.~Tabuchi, A.~Gloppe, K.~Usami, and Y.~Nakamura, Hybrid quantum systems based on magnonics, Appl. Phys. Express \textbf{12}, 070101 (2019).
    https://doi.org/10.7567/1882-0786/ab248d

\end{thebibliography}
\end{document}